\documentclass{jfm}

\usepackage{ulem}
\usepackage{natbib}
 \usepackage{colortbl}
\title{Stabilising Evaporating Soap Films with Salt}

\definecolor{darkgreen}{rgb}{0,0.5,0}
\definecolor{myblue}{RGB}{92, 92, 255}
\definecolor{mygreen}{RGB}{92, 255, 154}
\definecolor{myyellow}{RGB}{255, 226, 92}
\definecolor{myred}{RGB}{255, 102, 93}

\author{Victor Ziapkoff\aff{1}, François Boulogne\aff{1}, Anniina Salonen\aff{2},  \and Emmanuelle Rio\aff{1}}

\affiliation{\aff{1}Université Paris-Saclay, CNRS, Laboratoire de Physique des Solides, 91405, Orsay, France.
\aff{2}Soft Matter Sciences and Engineering, ESPCI Paris, PSL University, CNRS, Sorbonne Université, 75005 Paris, France}

\begin{document}
\maketitle

\begin{abstract}

We investigate the effect of a high concentration (32.5~g.L$^{-1}$) of sodium chloride (NaCl) on TTAB (tetradecyltrimethylammonium bromide) vertical soap films also called foam films, pulled out of a bath under controlled humidity conditions.
We observe that the film lifetime increases with relative humidity, both in the presence and absence of salt.
At any given humidity, the presence of NaCl systematically enhances film stability.
Our film thickness measurements show that the thinning dynamics with or without salt are nearly identical down to 100~nm.
Down to that thickness, the effect of evaporation can be rationalised by a constant evaporation rate, which becomes non-negligible compared to the drainage rate at film thicknesses below 400~nm.
The main effect of salt is the stabilisation of a Newton black film at a thickness of approximately 5~nm, whereas in the absence of salt, the film ruptures upon reaching a critical thickness of about 10~nm.

\end{abstract}

\begin{keywords}
Soap Film, Foam Film, Film lifetime, Thickness measurement, Evaporation rate
\end{keywords}

\section{Introduction}
\label{sec:Introduction}

Surface bubbles generated during coastal wave breaking eventually rupture. This leads to the generation of aerosols and microdroplets \citep{Deike2022}, which play a key role in ocean-atmosphere coupling.

The size of these ejecta is controlled, among other parameters, by the thickness of the fragile liquid film surrounding the bubble at the time of rupture. The latter depends on the thinning dynamics and on the lifetime of the film. A description of the dynamics of liquid films is therefore crucial to understand surface bubble stability.

Film thinning dynamics is driven by the coupled effects of drainage and evaporation, set by environmental conditions like atmospheric humidity \citep{champougny_influence_2018,auregan_drainage_2024,poulain2018ageing} and presence of convection \citep{auregan_drainage_2024}.

One commonly studied experimental configuration is vertical liquid films.
\citet{champougny_influence_2018} showed that, in this configuration, drainage dominates during the early stages of film thinning, when the film remains relatively thick.
As the film becomes thinner (on the order of one micrometer), evaporation becomes increasingly important in the dynamics.
Eventually, the film ruptures.

Additionally, natural convection and thermal effects \citep{boulogne2022measurement} must be taken into account, which complicates the derivation of comprehensive models.
Drainage is caused by the generation of thin patches in the vicinity of the meniscus, which rise by buoyancy \citep{seiwert2017velocity,miguet2021marginal}.
Recently, \citet{monier_self-similar_2024} proposed a description of the self-similar thickness profile of the thin films evolving under such a mechanism.
However, neither drainage nor evaporation rate are described quantitatively in the literature.

Another factor of complexity is the remarkable biological richness of the marine surface \citep{Helm2021}.
Additionally, the ionic composition of seawater is complex and variable as described by Dittmar’s law \citep{CopinMontegutSel}.
This makes it difficult to isolate the specific
influence of a single component on the behaviour of bubbles and films.

The effect of salts --- and more generally of electrolytes --- on adsorption/desorption of surfactant, water/air interfaces, and thin film structure has been extensively investigated in the scientific literature.
Regarding water/air interfaces, \citet{Onsager1934} demonstrated theoretically that increasing salt concentration leads to a rise in surface tension.
The effect has been later refined by Jones and Ray \citep{Jones1935_SurfaceTension,Jones1937_SurfaceTensionElectrolytes}, who observed little deviations at very low electrolyte concentrations.
This so-called Jones-Ray effect remains the subject of active debate \citep{Okur2017_JonesRayEffect}.
Another well-documented effect is the importance of electrolytes on the surfactant critical micelle concentration (CMC)
, which can be much smaller in presence of salt \citep{Roy2019,Qazi2020_DynamicSurfaceTension}.
The Krafft temperature, below which the salt precipitates, and the surfactant adsorption/desorption process are also modified in presence of electrolytes \citep{Chang1992,Roy2014,Roy2019}.

Early experimental work by \citet{ReinoldRucker1881,reinold_thickness_1893} revealed a decrease in the thickness of soap films as salt concentration increased.
This phenomenon has been theoretically rationalised by the models proposed by \citet{helmholtz_studien_1879,gouy_sur_1910,chapman_li_1913,Stern1924} and \citet{Debye1923}, who introduced the concept of electrostatic screening and elucidated how ionic strength modulates the spatial distribution of charges in aqueous solutions.
As a result, the addition of salt induces variation in electrostatic pressure.
This impacts the well-known DLVO theory \citep{derjaguin_theory_1941,verwey_theory_1948,Scheludko1959,Scheludko1960} that rationalises the balance between van der Waals attraction and electrostatic repulsion in the interaction between two interfaces.

Within the DLVO framework, two equilibrium states are distinguished for thin foam films: the Common Black Film (CBF), stabilised by electrostatic interactions, and the Newton Black Film (NBF), stabilised by steric repulsion \citep{IUPAC1971_DefinitionsColloid}.
The first ones have a thickness of few tens of nanometers, whereas the second ones have a thickness of about 5 nm \citep{schulze-schlarmann_disjoining_2006}.
The presence of salt decreases the repulsion between the interfaces, destabilising the CBF and densifies the surfactant monolayers, stabilising the NBF.
\citet{exerowa_common_1981} introduced the concept of a critical electrolyte concentration $C_{\mathrm{el,cr}}$, a threshold that governs the transition between the two stable configurations, passing from CBF to NBF.

More recently, the influence of electrolytes on the thinning of vertical foam films, which we will also call soap films in the following, has been reported on non-ionic surfactant systems. \citet{auregan_drainage_2024} demonstrated that the presence of salts does not modify drainage in any humidity conditions. Even at salt concentrations up to twice that of seawater, the only effect of salt addition on the drainage of thin films is through viscosity, and the thickness $h$ follows the scaling predicted by \citet{monier_self-similar_2024}.

However, the stability of liquid films is inherently a dynamic problem. Such films are continuously formed not only at the moment of bubble generation but also within foams during structural rearrangements when two bubbles come into contact, and a new film is created.
We thus propose to explore the stability of thin liquid films in a dynamic configuration.
\citet{saulnier_study_2014} developed a set-up that continuously generates vertical films, which allows to explore film formation under controlled conditions.
With this configuration, the soap film generated by pulling a frame out of a bath has an initial thickness of a few micrometers fixed by the Frankel law \citep{mysels_soap_1959,Mysels1962,van2008thickness}.
It thins over time as a result of capillary and gravity-driven drainage and evaporation and eventually ruptures.

In this article, we investigate the effect of a high concentration of salt, comparable to seawater, on film evaporation, drainage, and rupture during film generation. We present the experimental set-up for characterising vertical films in Section~\ref{sec:set-up} and the results obtained on film lifetime in Section~\ref{sec:FilmLifetime}. We rationalise the enhanced film stability in the presence of salt through measurements of their thickness. We demonstrate in Section~\ref{sec:FilmThickness} that salt enables the formation of a stable NBF \textit{i.e.} a constant thickness. The results are discussed in Section~\ref{sec:Discussion}.

\section{Experimental set-up}\label{sec:set-up}

\begin{center}
\begin{figure}
    \centering    \includegraphics[width=1\textwidth]{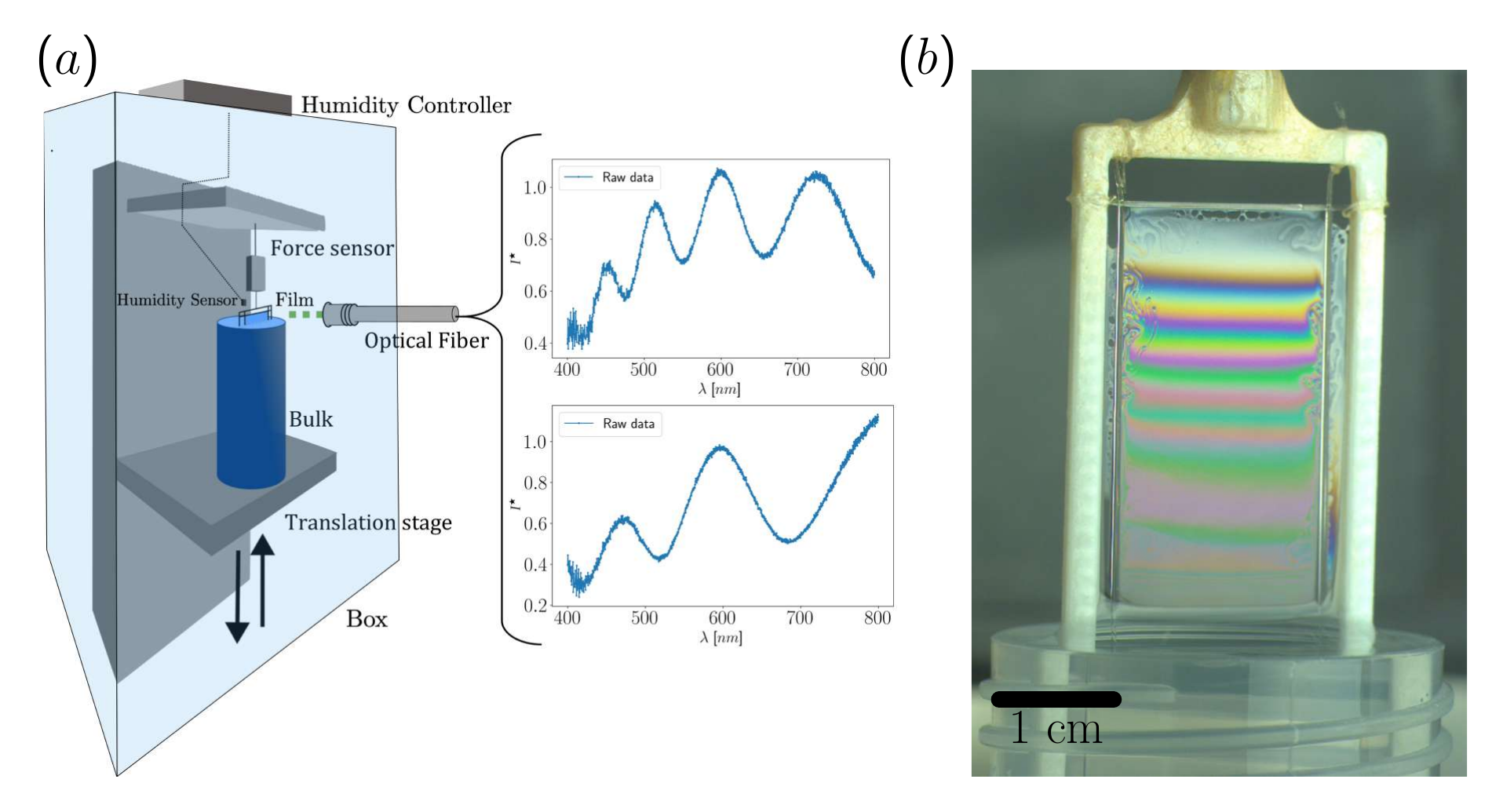}     \caption{\textit{($a$) Set-up scheme of a film generated by translating the reservoir downwards at constant velocity. The force sensor detects the film rupture, and the optical fiber measures the spectrum of the light reflected by the thin film close to the top. The figure exhibits two typical examples of spectra with wavelengths $\lambda \in$ [400, 800] nm. ($b$) Photograph of a foam film pulled out of a bath.}}\label{fig1}
\end{figure}
\end{center}

To characterise the thinning and rupture of vertical foam films during their generation, we use a setup developed by \citet{saulnier_study_2014}.
Fig. \ref{fig1} ($a$) shows a scheme of the setup.
The procedure to generate soap films consists in dipping a 3D-printed rectangular frame of surface area 10~cm~$\times$~1.7~cm in a circular reservoir of 2.8~cm diameter and 9.7~cm depth.

It contains an aqueous solution of tetradecyltrimethylammonium bromide (TTAB  - Sigma Aldrich) at the CMC, $C=$~1.18~g$.$L$^{-1}$.
We perform experiments with or without addition of sodium chloride (NaCl, Sigma Aldrich) at a concentration $C_s^\circ = $~32.5~g$.$L$^{-1}$.
Note that at this salt concentration the CMC decreases to $0.03 \ $g.L$^{-1}$, as expected (see Appendix \ref{appA}). However, we choose to keep the same concentration of TTAB even if the CMC is now lower.
To ensure sample purity, we recrystallise the TTAB \citep{Stubenrauch2005} and roast the salt overnight at 700~$^\circ$C.
A sub-frame, consisting of two vertical and one horizontal fishing wires, made with nylon fibre of diameter $150~\mu$m, is glued at 2~mm from the main frame.
The latter is connected to a force sensor (HBM, 5g) that detects film rupture.
The initial position of the manually adjustable translation plate is set so that the horizontal nylon fiber is nearly in contact with the bulk, which sets the zero vertical position.
To make the film, the translation plate moves downward at a constant velocity $V$ = 1~mm.s$^{-1}$ thanks to a motorised linear stage (Newport UTS150CC) coupled to a motion controller (Newport SMC100CC).
We set the time origin $t = 0$ of film generation to coincide with the start of the translation plate's movement. The maximum range of the translation plate is 81~mm, which limits our study to film lifetimes of 81~s.
Above this time, the film length remains constant.
Fig. \ref{fig1} ($b$) shows a photograph of a film during its generation.

As in \cite{champougny_influence_2018}, the film generation setup described above is enclosed in a box of dimensions 32~×~32~×~62 cm$^3$ in which humidity is regulated using a home-made controller described in \cite{boulogne_humidity_2019}.
A PID controller based on an Arduino Uno and a humidity sensor (Honeywell, HIH-4000-003) positioned near the top of the film allows the injection of appropriate proportions of dry and moist gas to achieve the target humidity in the box.
Dry gas is produced by circulating nitrogen directly in the box.
Moist air is obtained by bubbling air through water.
We perform experiments at four different relative humidities $\mathcal{R}_h  \in [40, 60, 80, 100]~\% $.
To achieve measurements at saturation $\textit{i.e.}$ at $\mathcal{R}_h$~=~100~$\%$, we saturate the atmosphere before starting regulation by sprinkling the edges of the box with water.
The whole setup yields a typical uncertainty of $\pm$~0.5~$\%$ on the relative humidity $\mathcal{R}_h$.
The temperature within the box is maintained at the room temperature, $\mathcal{T}$~=~20~$\pm$~1$^\circ$C.

A horizontal optical fiber (IDIL, France) is placed at the very top of the film, where the film ruptures, at the focal plane of the lens with a 400~$\mu$m spot size.
The reflected light is collected by six optical fibers (IDIL, France) surrounding the central one and its spectrum is measured by a spectrometer (NanoCalc 2000 VIS/NIR, Ocean Optics) in the wavelength range [400, 800]~nm with a time resolution of 50~ms.
This enables an acquisition throughout the full thinning process, \textit{i.e.} from film generation to rupture. Fig. \ref{fig1} ($a$) represents two spectra in the range [400, 800]~nm at two different times $t$.
It is clear from these examples that the signal-to-noise ratio is poor in the range [400, 450]~nm.
The range of wavelengths $\lambda$ studied is thus restricted to [450, 800]~nm.
The film thickness $h$ is then calculated from the spectra acquired manually and treated with the Python library \textit{optifik} relying on microinterferometry and presented in detail in {\citet{Ziapkoff2026WhiteLightInterferometry}}.
This method allows thickness measurements down to 5 with an associated uncertainty $\sigma$ = $\pm$~1~nm.

Film thickness measurements from interferometry require a precise determination of the refractive index.
Refractometer measurements (Abbemat MW, Anton Paar) have been performed on aqueous TTAB solutions at $C = 1.18$~g$.$L$^{-1}$ for wavelengths $\lambda = [481.3, 513.1, 589.3, 656.2]$~nm and temperature $\mathcal{T}$$\ = 20^\circ$ C, at six salt concentrations $C_s^\circ \in [0,10,20,30,40,50]$~g$.$L$^{-1}$.
This allowed us to determine \textit{ad hoc} Cauchy laws for our systems, $n(\lambda, C_s) = A + B/\lambda^2$ describing the index of refraction $n(\lambda)$.
The coefficients $A$ and $B$ for each condition are listed in Table \ref{tab:Cauchy}.
The coefficients $A = 1.329$ and $B = 3209$~nm$^2$  for $n(\lambda, C_s^\circ = 32.5~$~g$.$L$^{-1})$ are determined by interpolating the data in Table~\ref{tab:Cauchy}.
For clarity, in the following, the index of refraction without salt will be denoted as $n(\lambda, 0)$, and with salt as $n(\lambda, C_s^\circ)$.

\begin{table}
  \centering
  \begin{tabular}{ccc}
    $C_s^\circ$ [g.L$^{-1}$] & \textbf{$A$} & \textbf{$B$} [nm$^{2}$] \\
    0   & 1.324 & 3102 \\
    10  & 1.326 & 3136 \\
    20  & 1.327 & 3173 \\
    30  & 1.329 & 3201 \\
    40  & 1.331 & 3234 \\
    50  & 1.332 & 3270 \\
  \end{tabular}
  \caption{Coefficients $A$ and $B$ of \textit{ad hoc} Cauchy law, $n(\lambda,C_s^\circ) = A + B/\lambda^2$, determined at $\mathcal{T}$$\ = 20^\circ\ $C for aqueous TTAB solutions at $C = 1.18$ g.L$^{-1}$ with six different salt concentrations $C_s^\circ \in [0,10,20,30,40,50]$~g.L$^{-1}$. The refractometer measurements are done for wavelengths $\lambda = [481.3, 513.1, 589.3, 656.2]$~nm}\label{tab:Cauchy}
\end{table}

As soon as film rupture is detected by the force sensor, its lifetime $\tau_f$ is recorded, and a new one is generated.
This allows to accumulate statistics and to obtain histograms of film lifetimes $\tau_f$ with several hundreds of datapoints $N$.
Note that, as explained in the introduction, this setup is meant to measure foam film thinning and rupture during generation.
The choice of pulling velocity and frame length ensures that most films rupture during their generation.

\section{Film lifetime}\label{sec:FilmLifetime}

\begin{center}
\begin{figure}
    \centering
    \includegraphics[width=1\textwidth]{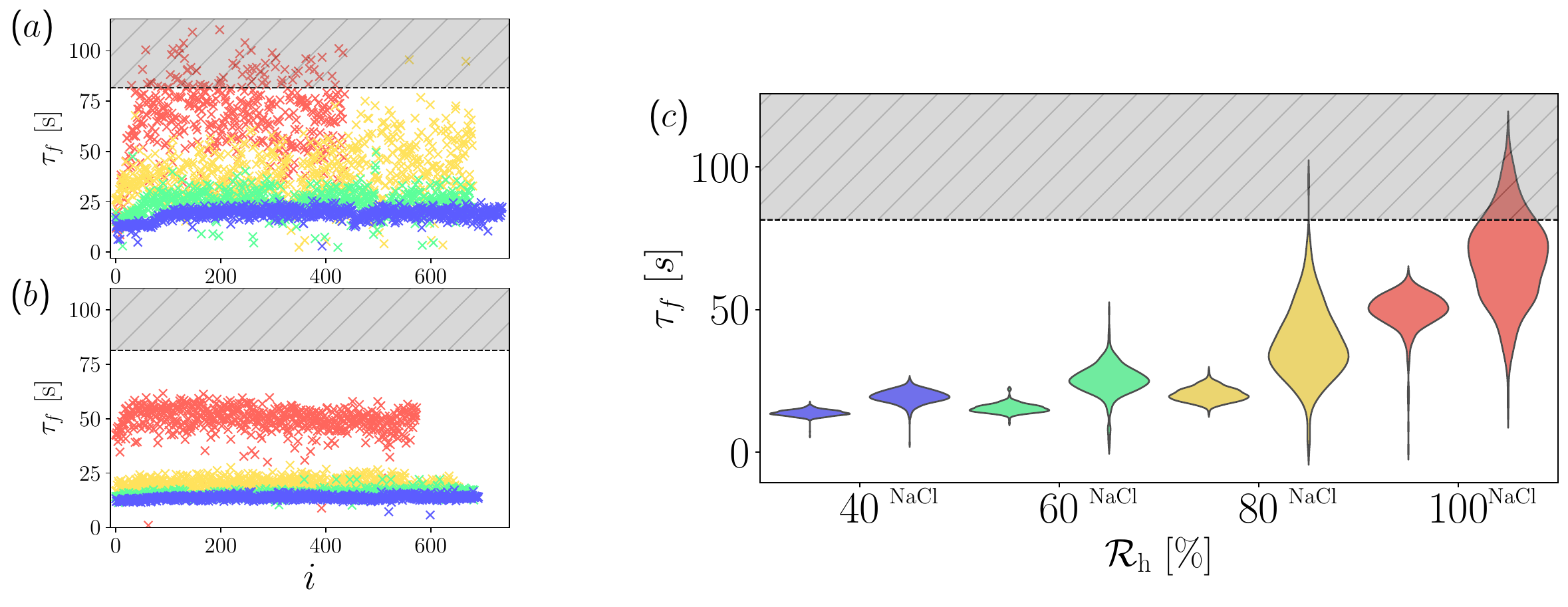}
    \caption{ \textit{Lifetime $\tau_f$ of TTAB films for relative humidities $\mathcal{R}_h  \in [40, 60, 80, 100]~\% $ represented in blue, green, yellow, red, respectively, for baths with ($a$) and without ($b$) the addition of 32.5~g$.$L$^{-1}$ NaCl. ($c$) Violin plots of the film lifetime distributions, using the same colour code. The label 'NaCl' indicates experiments performed with salt. The shaded region corresponds to lifetimes where the translation stage has stopped before bursting.}}\label{fig2}
\end{figure}
\end{center}

We create $N > 400$ films at four different relative humidities with and without sodium chloride in the bath.
Fig.~\ref{fig2}~$(a,b)$ presents the raw data of the lifetimes $\tau_f$ measured for each humidity level - displayed in blue, green, yellow, and red for $\mathcal{R}_h = 40\ \%, 60\ \%, 80\ \%,$ and $100\ \%$, respectively, with ($a$) and without ($b$) NaCl.
For $N < 100$, a transient regime is visible, likely due to the initial adsorption dynamics of the surfactant at both the bath and film interfaces.
In the following, those data are not considered in the analysis.
Fig.~\ref{fig2}~($c$) displays violin plots for all experimental conditions, using the same colour code. The NaCl labels at the bottom of the violin plots indicate the datasets corresponding to the saline solutions.
The shaded regions in all plots represent domains where the translation stage has reached its maximum range, and the film is no longer under generation.

From Fig.~\ref{fig2}, we observe that increasing ambient humidity leads to longer film lifetimes $\tau_f$, regardless of the presence of salt - an intuitive effect already reported by \citep{champougny_influence_2018,auregan_drainage_2024,poulain2018ageing}.
This trend is particularly pronounced in the salt-free system, where a clear jump in lifetime is observed between $\mathcal{R}_h  = 100 \ \%$ and lower humidity levels.
In contrast, the NaCl-containing systems show a more gradual and regular increase in film lifetime with humidity.
The presence of NaCl not only increases the mean film lifetime but also results in more dispersed distributions, as indicated by the longer violin plots (Fig.~\ref{fig2}~($c$)).

\section{Film thickness}\label{sec:FilmThickness}

In this section, we first address the impact of the relative humidity of air on film thinning, and subsequently quantify the effect of salt addition.

\subsection{Impact of the relative humidity of air }\label{subsec:humidityImpact}

To rationalise the increase in film stability induced by the presence of salt, we make 10 films at the four relative humidities studied and measure their thicknesses.
We observe that all thinning curves collapse, proving good reproducibility.
This motivates the use of only the longest film from each experimental condition for detailed analysis.

\begin{center}
\begin{figure}[h]
    \centering
    \includegraphics[width=1\textwidth]{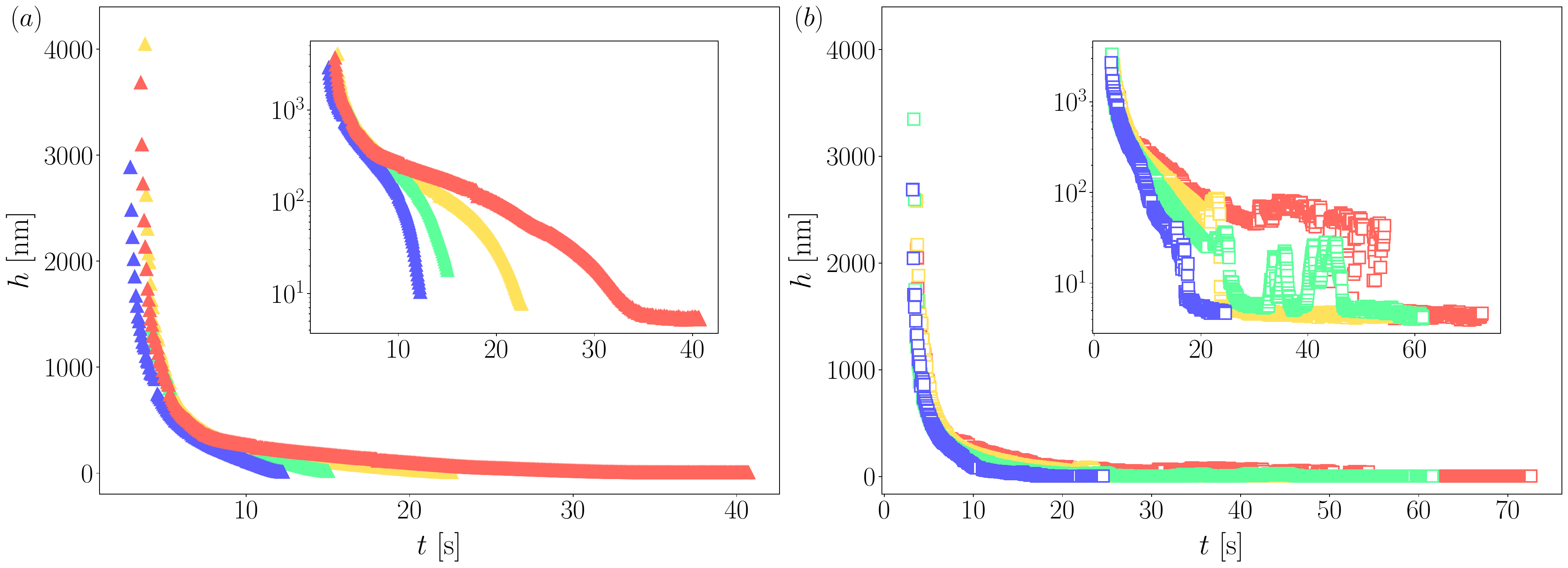}
    \caption{\textit{Time evolution of film thickness for the most stable TTAB film at each relative humidity tested ($\mathcal{R}_h = 40~\%, 60~\%, 80~\%, 100~\%$; blue, green, yellow, red), ($a$) without ($\blacktriangle$) salt and ($b$) with ($\square$) 32.5~g$.$L$^{-1}$ NaCl. Insets show the data in log-lin scale.}}\label{fig3}
\end{figure}
\end{center}

Fig.~\ref{fig3}~($a$) shows the evolution of the film thickness corresponding to the most stable films measured at each relative humidity studied in the salt-free system.
We observe that, aside from a slight horizontal shift of the curve at $\mathcal{R}_h = 40~\%$ (blue) likely due to an offset in the zero reference of the translation stage, all curves closely follow the same dynamics down to a thickness of 400~nm.
As shown in the inset of Fig.~\ref{fig3}~($a$) (log-lin scale), the curves start to deviate significantly from one another below 400~nm. We observe that, at such small thicknesses, a higher relative humidity leads to a slower thinning rate of the film.
As a consequence, at a given generation time $t$, films formed at higher humidity are systematically thicker.

This observation is in good agreement with the previous work by \citet{champougny_influence_2018}.
The authors considered that drainage and evaporation are decorrelated.
Under saturated vapor, where thinning is governed exclusively by drainage, \citet{champougny_influence_2018} observed that the drainage rate decreases over time. Under non-saturated conditions ($\mathcal{R}_h <$ 100~$\%$), evaporation also contributes to thinning, but its effect becomes significant only for film thicknesses under $h=400$~nm.
This does not necessarily mean that evaporation dominates over drainage in this regime, but that we need to consider both.
Eventually, the films rupture as soon as their thickness reaches 10~nm, indicating that no stable CBF or NBF form under these conditions.

Fig.~\ref{fig3}~($b$) shows four thinning curves corresponding to the most stable films at each tested relative humidity, for the system with added salt. As in the salt-free case, all four curves collapse onto a common curve down to 400~nm.
Below this thickness, the curves begin to differ significantly.
Thus, the drainage-dominated regime above 400~nm is followed by a regime below where evaporation plays an important role.
The inset in Fig.~\ref{fig3}~($b$) (log-lin scale) highlights the [$0, 400$]~nm range, where we observe local thickness variations.
These are likely due to the emergence of localised thick patches - an effect previously reported by \citet{auregan_drainage_2024}.
In Fig.~\ref{fig3}~($b$), we also observe the presence of a thickness plateau at all the studied humidities.
The corresponding thickness on this plateau is constant at approximately 5~nm\ independent on the relative humidity.
The higher the humidity, the longer the plateau.
We identify this plateau as corresponding to the formation of a stable Newton Black Film (NBF).
Indeed, in the literature, it is demonstrated that such films, stabilised by a steric repulsion between the surfactant heads at the interfaces, have a thickness between 5 and 10 nm \citep{schulze-schlarmann_disjoining_2006}.

The thickness at the plateau $h_p = 5$ nm is measured by microinterferometry using a single-layer model with two reflecting interfaces, analogous to a Fabry-Perot cavity \citep{Ziapkoff2026WhiteLightInterferometry}.
This model does not account for the surfactant layer.
However, at such thicknesses, the contribution of the surfactant layer cannot be neglected.
We can use multilayer models instead to refine the description of the film layer \citep{mysels_soap_1959}.
The three-layer model introduced by \citet{duyvis_phd_1962}  considers an aqueous core of thickness $h^\dagger$ including an ammonium TTAB head sandwiched between tetradecane tails of thickness $h_1$ and yields a total layer thickness $h_\mathrm{tot}$.
According to this model, the calculation of $h_\mathrm{tot}$ and $h^\dagger$ are given by

\begin{equation}
h_\mathrm{tot} = 2h_1 + h^\dagger; \quad h^\dagger= h_p - 2h_1 \frac{n_1^2 - 1}{n_2^2 - 1},
\end{equation}

\noindent where $n_1 = 1.4290$ is the refractive index of pure tetradecane \citep{lide_handbook_1993} and $n_2$ the refractive index of the aqueous core, assumed here as $n_2 = \frac{1}{m} \sum_{i=1}^{m} n(\lambda_i,C_s^\circ)$ with $m = 1106$ the number of discrete wavelength values in the experimental spectrum where $\lambda\in[450,800]$~nm.
The linear tetradecane chains oriented perpendicular to the interface yield a thickness $h_1$ of 1.7~nm \citep{simister_structure_1992}. Using these values, the actual thickness of the core at any given relative humidity $\mathcal{R}_h$ is $h^\dagger \approx 0.52$ nm, consistent with neutron scattering measurements reported by \cite{simister_structure_1992}.
Note that we have neglected the variation of the refractive index with the salt concentration.
We estimate an increase of $n_2$ of about 3 \% at saturation (from \citet{Li2015Optical}), in which experiments are performed on salty water, in the absence of surfactant).
Taking this into account would lead to $h^\dagger =1.13\ $nm.
Note that this error, which increases over time, will also impact the values of $h$ in Figure \ref{fig3} (b) by at most 3 \%.

\subsection{Impact of salt}\label{subsec:humidityImpact}

\begin{center}
\begin{figure}[h]
    \centering
    \includegraphics[width=1\textwidth]{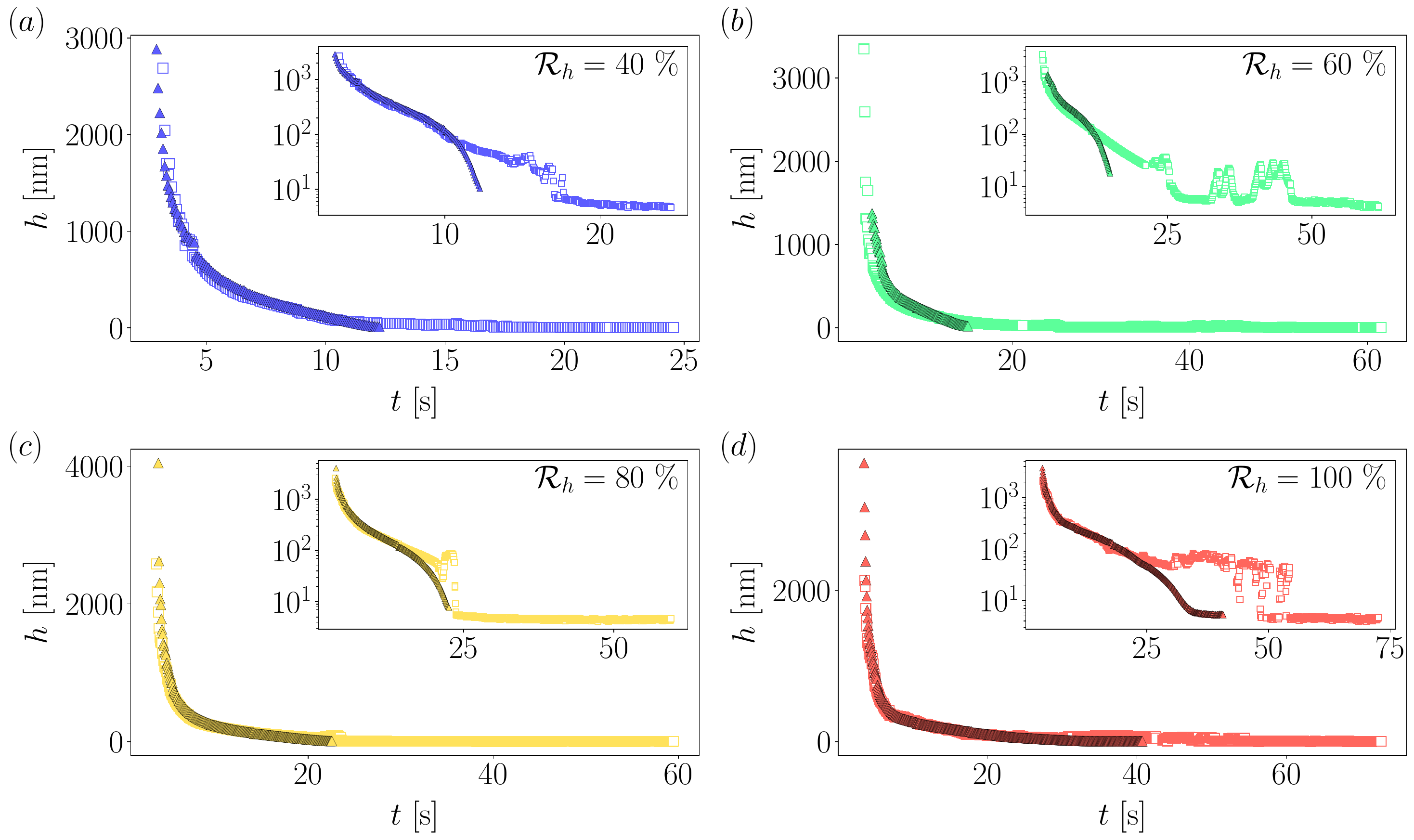}
    \caption{\textit{Thinning curves of films with ($\square$) and without ($\blacktriangle$) 32.5~g$.$L$^{-1}$ NaCl for the four fixed relative humidities: ($a$)~$\mathcal{R}_h = 40~\%$, ($b$)~$\mathcal{R}_h =60~\%$, ($c$)~$\mathcal{R}_h =80~\%$, and ($d$)~$\mathcal{R}_h =100~\%$. The $\blacktriangle$ markers for films without salt are outlined in black for clarity.
    Insets show the data in log-lin scale.}}\label{fig4}
\end{figure}
\end{center}

From the thinning curves presented in Fig.~\ref{fig3}, a direct comparison can be made for the liquid composition at fixed relative humidity~$\mathcal{R}_h$.
Fig.~\ref{fig4} shows the thinning dynamics with and without salt at $\mathcal{R}_h = 40~\%$, 60~\%, 80~\%, and 100~\% in subfigures ($a$), ($b$), ($c$), and ($d$), respectively.
Each figure includes an inset representing the data in log-lin scale.

In Fig.~\ref{fig4}, for each tested humidity, the thinning curves with or without salt follow the same dynamics
down to 100~nm in agreement with results reported previously by \cite{auregan_drainage_2024}.
In contrast, when $h < 100$~nm, the presence of salt tends to slow down thinning and to stabilise the NBF.

\section{Interpretation}\label{sec:Discussion}

In this section, we first address the stochastic nature of film lifetimes.
We then provide a detailed discussion of the thinning dynamics, followed by a quantification of the contribution of evaporation to film rupture.

\subsection{Film rupture stochasticity}

The stochastic nature of film rupture has been explored in different contexts, including public studies of horizontal film lifetimes sealed in bamboo cylinders by \cite{Tobin2011}, experimental studies on two-dimensional foams by \cite{forel_coalescence_2019}, droplet coalescence simulations by \cite{perumanath_droplet_2019}, and more recently, thin-film pressure balance (TFPB) measurements by \cite{chatzigiannakis_breakup_2020}.
However, the dynamics of vertical film rupture remains underexplored.

Given the data shown in Sec.~\ref{sec:FilmLifetime}, we quantify the degree of stochasticity in vertical film rupture by following the approach of \cite{shaw_dripping_1984}.
Specifically, we represent the difference of lifetime of consecutive films through return maps, plotting the interval

\begin{equation}
T_{i+1} = \tau_{f}^{i+2} - \tau_{f}^{i+1}
\end{equation}

\noindent between the lifetimes of films $i+2$ and $i+1$ as a function of the interval $T_{i}$ between the lifetimes of films $i+1$ and $i$.
If rupture events, considered here as independent, are strongly correlated, characteristic patterns should emerge in these maps.
Conversely, if rupture events are independent, the resulting cloud of points should be homogeneously distributed and centered, reflecting stochastic dynamics.

\begin{center}
\begin{figure}[h]
    \centering
    \begin{minipage}{0.5\textwidth} \includegraphics[width=\textwidth]{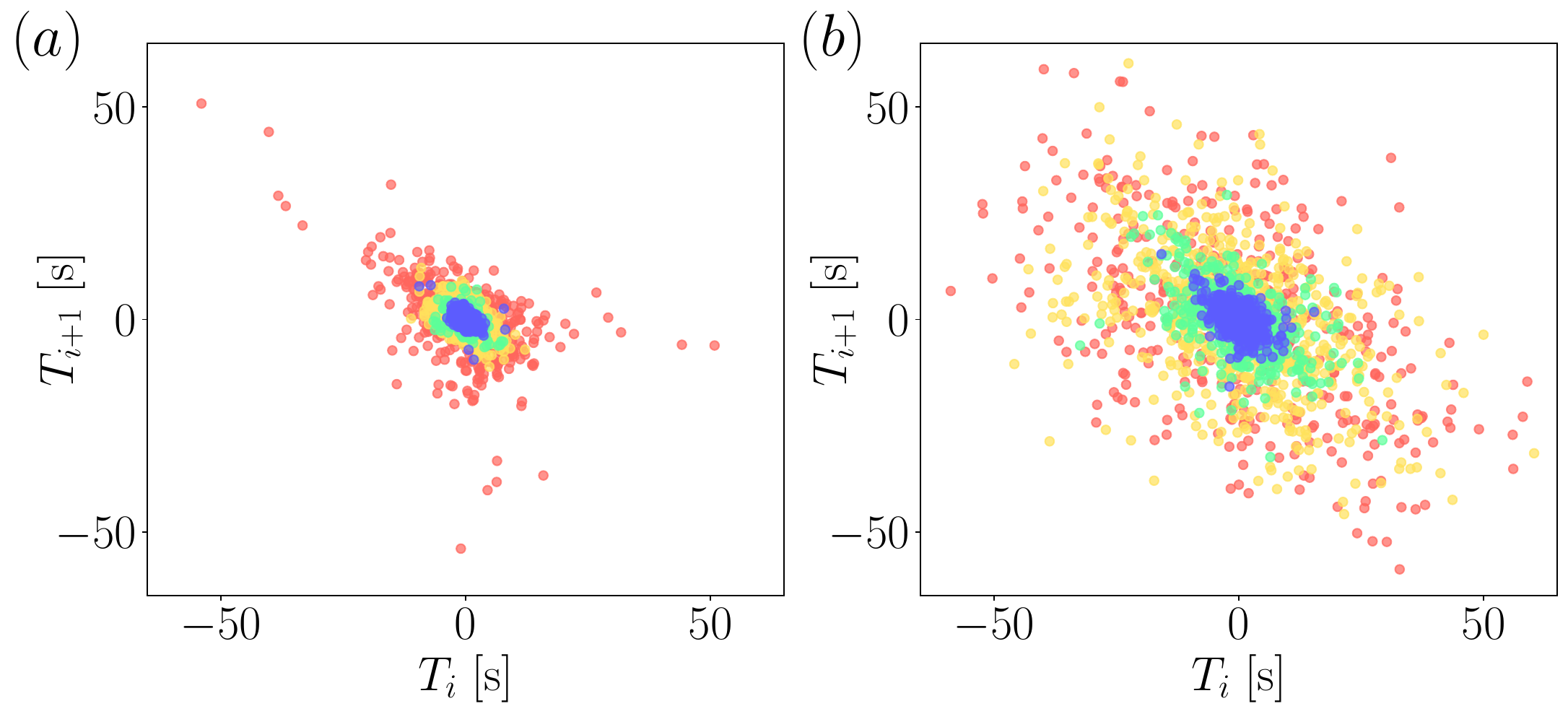}
        \centering
    \end{minipage}%
    \begin{minipage}{0.25\textwidth}
        \includegraphics[width=\textwidth]{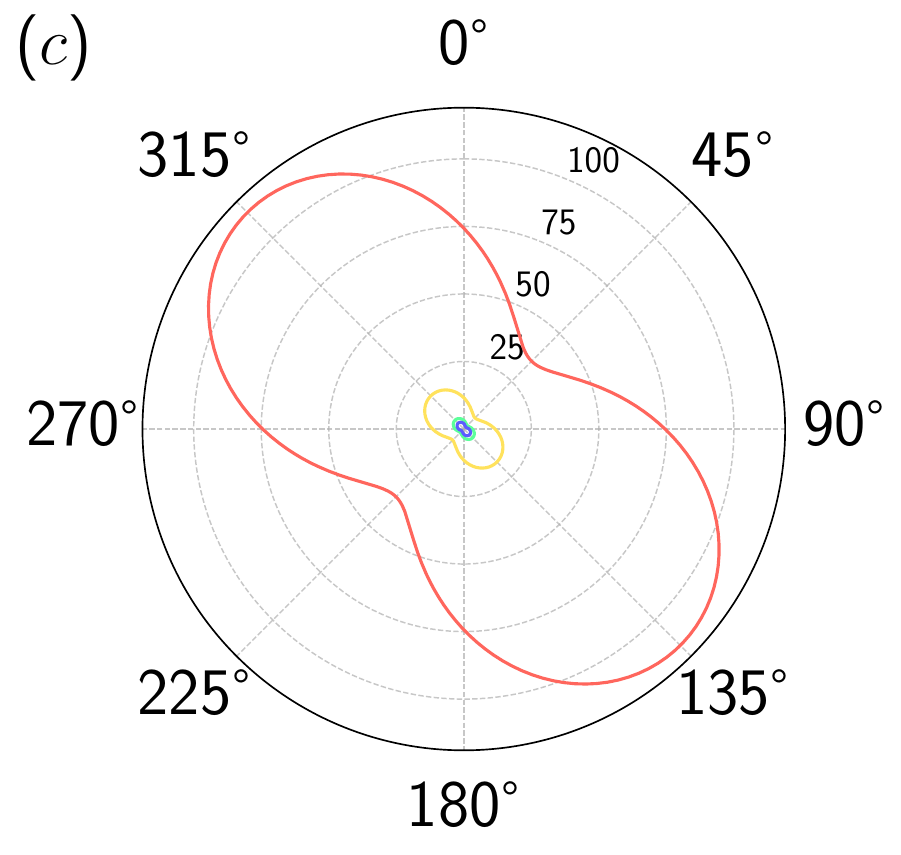}
        \centering
    \end{minipage}%
    \begin{minipage}{0.25\textwidth}
        \includegraphics[width=\textwidth]{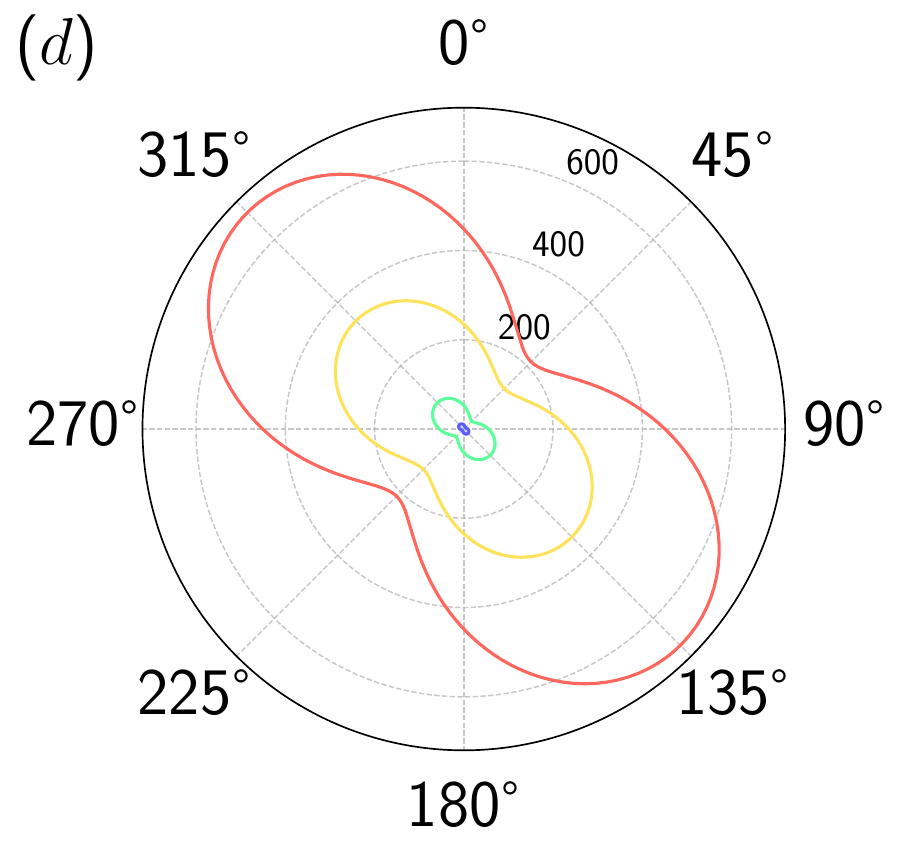}
        \centering
    \end{minipage}
    \caption{\textit{($a$)  Experimental return maps for films without salt at four relative humidities $\mathcal{R}_h \in [40, 60, 80, 100]\ \%$. ($b$) Experimental return maps for films with 32.5~g$.$L$^{-1}$ sodium chloride addition under identical humidity conditions. Polar representation of the variance $\langle z_i(\theta)^2\rangle$ as a function of the angle $\theta$ for films (c) without salt and (d) with salt addition, at relative humidities $\mathcal{R}_h \in [40, 60, 80, 100]\ \%$. All figures use the same colour code than in Fig.~\ref{fig2}.}}\label{fig6}
\end{figure}
\end{center}

Figures~\ref{fig6}~($a$) and \ref{fig6}~($b$) display return maps for films with or without salt at four different relative humidities, $\mathcal{R}_h \in [40,60,80,100] \ \%$, using the same colour code as in Fig.~\ref{fig2}.
To quantify any privileged angular directions from the cloud data points, we construct a new variable $z_i(\theta)$ defined as
\begin{equation}
    z_i(\theta) = \left(T_{i} - \tilde{T}\right )\cos\theta
                + \left(T_{i+1} - \tilde{T}\right)\sin\theta,
                \label{eq:z_i}
\end{equation}

\noindent with $i\in[0, N-1]$ and $\theta \in [0,2\pi]$ and where the mean value is defined as

\begin{equation}
\tilde{T} = \frac{1}{N}\sum_{i=1}^{N} T_{i}.
\label{T_N}
\end{equation}

Figures~\ref{fig6}~($c$) and \ref{fig6}~($d$) then present the variance
\begin{equation}
\langle z(\theta)^2\rangle = \frac{1}{N}\sum_{i=1}^{N} z_i(\theta)^2,
\end{equation}

\noindent plotted in polar coordinates as a function of $\theta$.
For clarity, angles are expressed in degrees in these polar representations.

The results in Fig.~\ref{fig6}~$(c,d)$ indicate that, for both systems and at all relative humidities, the variance is consistently more pronounced along the $315^{\circ}$--$135^{\circ}$ direction, with the amplitude of the variance lobes increasing with humidity.
This trend reflects a negative correlation between rupture events.
When $T_{i}$ is larger than average, $T_{i+1}$ tends to be smaller, and $\textit{vice versa}$.
In other words, long films are statistically more likely to be followed by shorter ones.
Salt addition does not significantly alter this behavior.
This robust feature suggests that film rupture is neither purely stochastic nor fully deterministic, but rather exhibits intermediate dynamics.

\subsection{Film evaporation without salt}\label{subsec:evapSansSel}

In this section, we provide a quantitative description of the film thinning dynamics over the entire film lifetime, under all humidity conditions, to rationalise the results reported in Sec.~\ref{sec:FilmLifetime}.
We consider here only salt-free films.
As shown in Fig.~\ref{fig3}~$(a)$, small offsets are present in experimental data, which we attribute to slight differences in the initial positioning of the translation stage.
We correct this artifact by shifting each data set in time, as explained in the following.

\begin{center}
\begin{figure}[h]
    \centering    \includegraphics[width=0.75\textwidth]{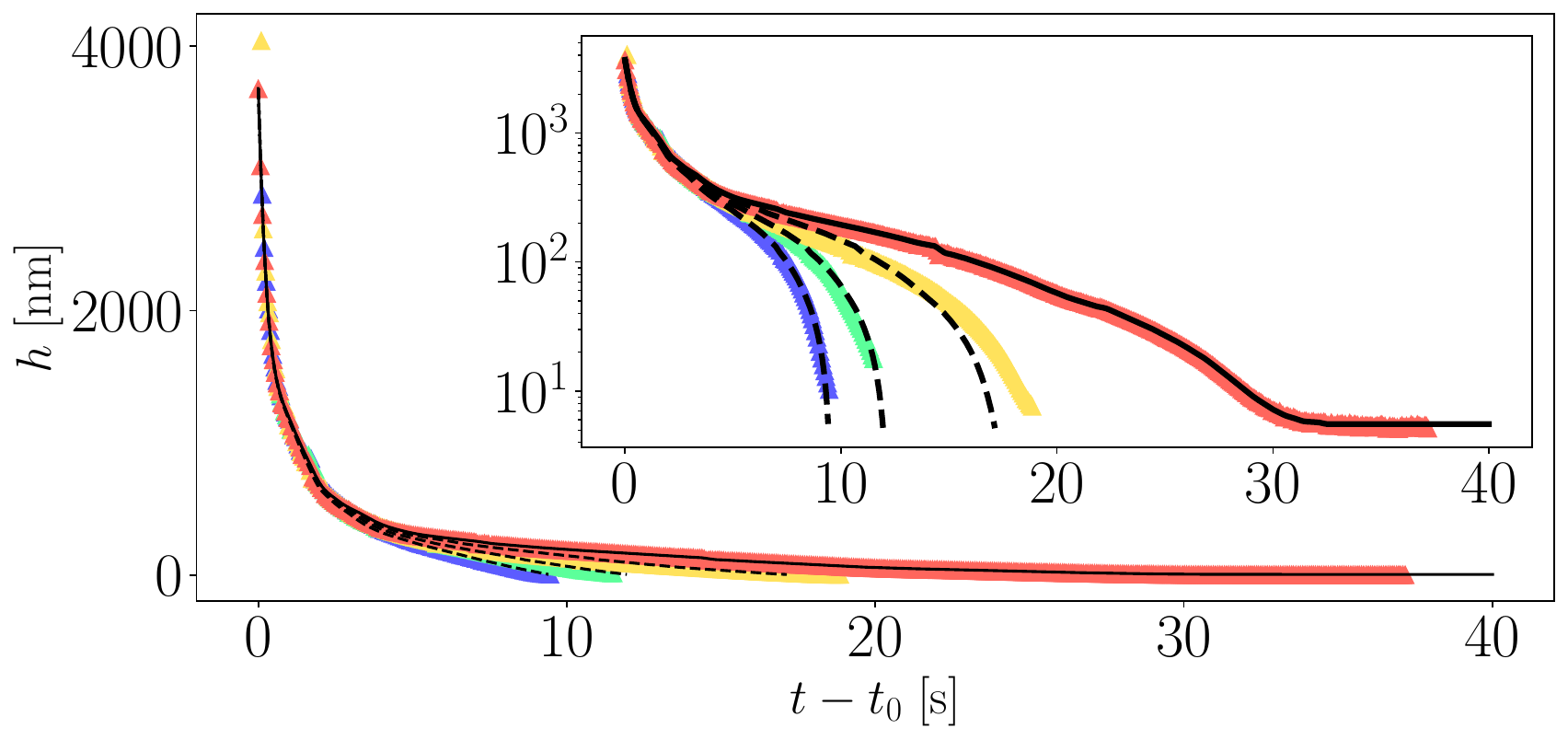}
    \caption{\textit{Time shifted evolution of film thickness for the TTAB film showed in Fig.~\ref{fig3}~$(a)$ at relative humidity $\mathcal{R}_h = 40~\%, 60~\%, 80~\%, 100~\%$ without salt. The respective time shifts $t_0$ at $\mathcal{R}_h = 40~\%, 60~\%, 80~\%$ are 0.764~s, 0.047~s, and 0.166~s. The black line represents the computation of Eq.~\ref{eq:amincissement} at saturation, whereas dashed lines represent the computation at relative humidity $\mathcal{R}_h = 40~\%, 60~\%, 80~\%$. The fitting values of the prefactor $a$ are $\{0.74,0.67,0.55\}$ and the evaporation rate $j_{\mathrm{e}}$ is $[21, 13, 5] \ $nm.s$^{-1}$, for the three different humidities. Insets show the models and experimental data in log-lin scale.}}
    \label{fig8}
\end{figure}
\end{center}

We observe that the early stages of film thinning are mostly governed by drainage and not depend on humidity.
This leads us to take the experimental data obtained in a saturated environment, displayed as red triangles, as a reference.
We minimise the least-squares error between the $t(h)$ curves and the reference curve, restricted to the interval $h \in [1000, 3000]$~nm, in the early stage of the film dynamics and with a sufficient number of data points to determine the optimal temporal shifts $t_0$. The corresponding temporal corrections are smaller than one second.
The resulting shifted datasets are shown in Fig.~\ref{fig8}, using triangular symbols and the same colour code as in the previous sections.

To determine theoretically the evolution of the thickness in time $h(t)$, we use the approach by \citet{pasquet_lifetime_2024} and by \citet{champougny_influence_2018}.
In the following, we make the hypothesis that the film thickness is uniform.
The slope of each curve in Fig.~\ref{fig8} writes,

\begin{equation}
\left.\frac{\rm{d} h}{\rm{d} t}\right|_{\mathcal{R}_h,\, \mathcal{T}}
= -\,2 j_{\mathrm{e}} - j_{\mathrm{d}},
\label{eq:amincissement}
\end{equation}

\noindent where $\mathcal{T}$ is the film temperature, assumed to be equal to the room temperature, $j_{\mathrm{e}}$ is the water evaporation rate from the film, the factor 2 comes from the presence of two interfaces, and $j_{\mathrm{d}}$ is the drainage rate inside the film, supposed independent of the atmospheric humidity.
Both contributions are defined as positive.
By definition, in the saturated experimental data, the evaporation rate $j_{\mathrm{e}} = 0$ and
$\left.\frac{\rm{d} h}{\rm{d}  t}\right|_{\mathcal{R}_h = 100\ \%,\, \mathcal{T}} =  - j_{\mathrm{d}}$.

To extract the drainage curves $j_{\mathrm{d}}(h)$
, we first smooth the experimental data obtained in a saturated atmosphere with a third order Savitzky–Golay filter \citep{savitzky1964smoothing} with a window size of 11 points and derive the temporal data $h(t)$.
The derivation, noted $j_{\mathrm{d}}(t)$, is smoothed with the same procedure, and an interpolation 
gives us the drainage rate $j_{\mathrm{d}}(h)$.
We numerically solve Eq.~\ref{eq:amincissement} with the function \textit{odeint} from the \textit{scipy.integrate} package and the result is displayed in Fig.~\ref{fig8} as a black line.
We restrain our computation to $h>5$ nm because we do not take into account disjoining pressure, important at these length-scales.
As expected, the numerical solution accurately follows the experimental data.

To perform this analysis for data at any relative humidity $\mathcal{R}_h$, following Eq.~\ref{eq:amincissement}, we need to determine the evaporation rate $j_{\mathrm{e}}(h)$ for any thickness.
We use the framework established by \citet{SchmidtBeckmann1930}, \citet{Dollet2017},   \citet{Boulogne2018}, and \citet{pasquet_lifetime_2024} to obtain an expression for $j_{\mathrm{e}}(h)$.
In particular, \cite{Boulogne2018} demonstrated that natural convection evaporation significantly affects vertical aqueous films of a few centimeters or more, a criterion that we meet (Sec.~\ref{sec:FilmLifetime}).
This observation is also supported by the direct calculation of the Grashof number,
\begin{equation}
\mathrm{Gr} = \frac{g}{\nu^2} \frac{\rho_\infty - \rho_s}{\rho_\infty} \mathcal{L}^3,
\label{eq:def_grashoff}
\end{equation}

\noindent where $g = 9.81$~m.s$^{-2}$ is the gravitational acceleration, $\nu = 1.5 \times 10^{-5}$~m$^{2}$.s$^{-1}$ the kinematic viscosity of water vapor in air, $\rho_s$ and $\rho_\infty = \rho_s\mathcal{R}_h$, the air densities at saturation and far from the evaporating surface, respectively, and $\mathcal{L}$ the characteristic vertical length scale of the evaporating film, which we take equal to $20$ mm for the calculation.
$\mathrm{Gr}$ quantifies the relative importance of buoyancy-driven flow.
Large $\mathrm{Gr}$ values indicate that natural convection dominates the evaporative transport.
Following \cite{Jones1978} and \cite{Boulogne2018}, we compute $\rho_s$ and $\rho_\infty$ at any relative humidity.
Across all experimental conditions, we consistently find $\mathrm{Gr} \gg 1$ (typically $\mathrm{Gr} > 200$), confirming that buoyancy-driven convection is the dominant mechanism.

Accordingly, the evaporative rate of a film can be approximated as
\begin{equation}
j_{\mathrm{e}} = a \frac{ \hat{c}_0 \mathcal{D}\left(c_{\mathrm{sat}}-c_{\infty}\right)}{\rho_w} \frac{\mathrm{Gr}^{1 / 4}}{\mathcal{L}},\label{eq:je_model}
\end{equation}
\\
\noindent where $\hat{c}_0 = 0.478$ is a prefactor arising from a numerical solution \citep{Boulogne2018},  $\mathcal{D} = 2.44 \times 10^{-5}$~m$^{2}$.s$^{-1}$ is the diffusion coefficient of water vapor in air, $c_\mathrm{sat}$ and $c_\infty =c_\mathrm{sat}\mathcal{R}_h$ are the mass concentrations of saturated vapor and ambient water vapor, respectively and $\rho_w = 1000$~kg.m$^{-3}$ is the water density. At $\mathcal{T} = 20^{\ \circ}$C, $c_\mathrm{sat} \approx 17$~g.m$^{-3}$ for water.
The prefactor $a$ will be used as a fit parameter.

We solve Eq. \ref{eq:amincissement} \cite{pasquet_lifetime_2024} using the explicit expression of $j_e$ \citep{Boulogne2018} given by Eq. \ref{eq:je_model}.
We start all computations at $t=t_0$ and $h=3.2\ \mu $m, which is the first thickness detected in the reference data.
We numerically solve Eq.~\ref{eq:amincissement}  with the function \textit{odeint} from the \textit{scipy.integrate} package.
The result is displayed in Fig.~\ref{fig8} as  dashed black lines for relative humidities $\mathcal{R}_h \in [40, 60, 80]~\%$ and captures the experimental thinning dynamics.
Indeed, the values of $j_{\mathrm{e}} \in [21, 13, 5]\,\mathrm{nm .s^{-1}}$ are comparable with the values $j_{\mathrm{e}} \in  [16, 8, 3.5]\,\mathrm{nm . s^{-1}}$ reported by \citet{champougny_influence_2018}.

The fitting parameter $a$ accounts for deviations from the idealised model, such as convection, thermal exchange, complex geometry (including the presence of the bath), the variation of $\mathcal{L}$ and Gr in time, and the presence of concentration and temperature gradients.
Notably, $a$ (see Table \ref{tab:Summary_a}) remains of order 1, indicating that our expression of $j_e$ captures the dominant physical mechanisms.

\begin{table}
  \centering
  \begin{tabular}{|c||c|c|c|c|}
    $\mathcal{R}_h$ [$\%$] & 40 & 60 & 80 &100\\
    $a$   & \cellcolor{blue} 0.74
    & \cellcolor{green} 0.67
    & \cellcolor{yellow} 0.55
    & \cellcolor{red} -- \\
    $a'$
    & \cellcolor{blue} 1
    & \cellcolor{green} 0.7
    & \cellcolor{yellow} 0.7
    & \cellcolor{red} 0.7\\

  \end{tabular}
  \caption{Summary of the fitted values of parameters $a$ (Eq. \ref{eq:je_model}) and $a'$ (Eq. \ref{eq:je_model_avec_sel}). }\label{tab:Summary_a}
\end{table}

\subsection{Film evaporation with salt}

\begin{center}
\begin{figure}[h]
    \includegraphics[width=1\textwidth]{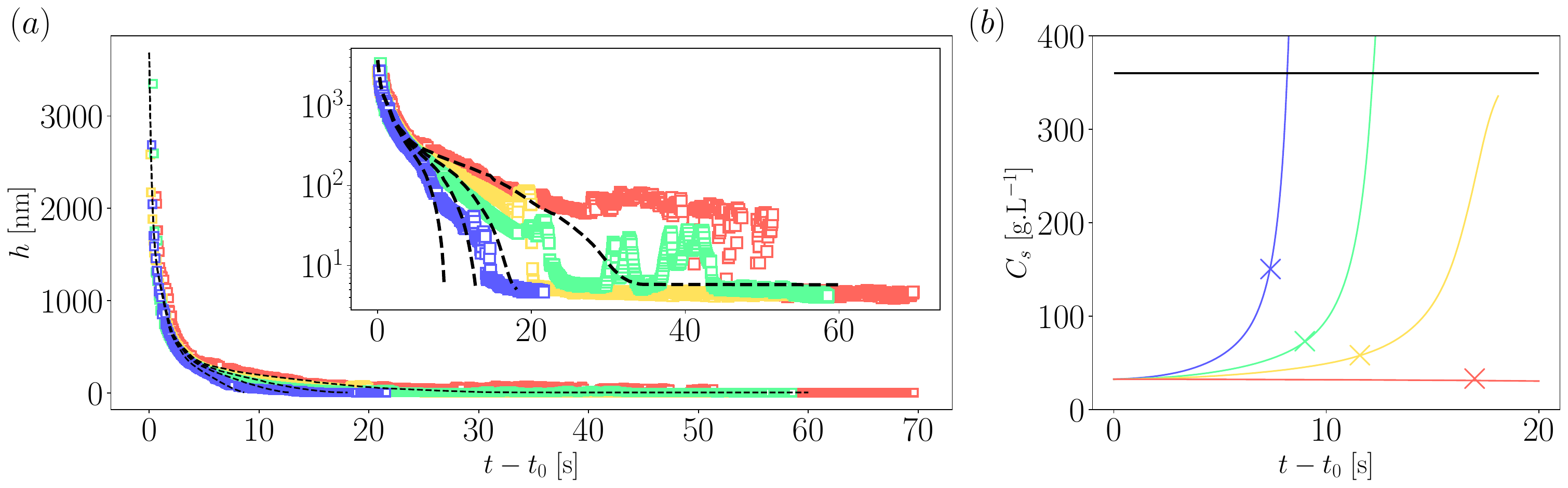}
    \caption{\textit{$(a)$ Time shifted evolution of the film thicknesses for TTAB films with salt shown in Fig.~\ref{fig3}~$(b)$.
    The dashed lines represent the computation of $h(t)$ at the four relative humidities with the function \textit{odeint} for Eq.~\ref{eq:amincissement} and Eq.~\ref{eq:evol_Salt}.
    The respective time shifts $t_0$ are 0.672~s, 0.669~s, 0.105~s and 0.672~s and the values of the prefactor $a'$ are $\{1,0.7,0.7,0.7\}$.
    Insets show the models and experimental data in log-lin scale.
    $(b)$ Numerical solutions of the salt concentration from Eq.~\ref{eq:amincissement} and Eq.~\ref{eq:evol_Salt} for salt concentration $C_s$ with relative humidity $\mathcal{R}_h = 40~\%, 60~\%, 80~\%, 100~\%$. Coloured crosses correspond to $t^\star$ such as $h^\star(t^\star) = 100$ nm, where the corresponding salt concentrations are $C_s(t^\star)$. The black straight line represents the saturation of salt in water.
    }}\label{fig9}
\end{figure}
\end{center}

To rationalise the results presented in Fig.~\ref{fig3}~$(b)$, we follow the same approach as in Subsec.~\ref{subsec:evapSansSel}.
As observed in Fig.~\ref{fig4}, salt has no significant effect on drainage.
We also previously argued that drainage also does not depend on humidity.
Consequently, to mitigate potential offset on experimental data, we shift by $t_0$ all datasets in time with respect to the same reference -- the red triangle displayed in Fig.~\ref{fig8} -- using the same least-squares error minimisation than in Subsec. \ref{subsec:evapSansSel}.
The corresponding temporal corrections for all humidity conditions tested are smaller than one second. Fig.~\ref{fig9}~$(a)$ shows the shifted datasets using square symbols and the same colour as in the previous sections.
The inset is the log-lin representation.

The difficulty in applying the model developed in Subsec.~\ref{subsec:evapSansSel} arises because, as the film thins by evaporation, the salt concentration $C_s$ increases and, in turn, affects the evaporation described in Eq. \ref{eq:je_model}.
This coupled effect results from the dependence of the parameters $c_\mathrm{sat}(C_s)$ and $\mathrm{Gr}(C_s)$ on the salt concentration.
Additionally, it is well established in the literature \citep{OBrien1948,kou_vapor_1999, Generous2020,lopez-borrell_vapor_2024} that the presence of salt affects the humidity fixed point.
For instance, at $C_s = 32.5\ \mathrm{g.L^{-1}}$, the liquid-vapor equilibrium is reached at $\mathcal{R}_h = 98\ \%$, while at saturation ($C_s = 360\ \mathrm{g.L^{-1}}$ at $\mathcal{T} = 20^\circ \ \mathrm{C}$), the equilibrium humidity drops to $\mathcal{R}_h \approx 78\ \%$.
A full modeling of the film dynamics requires an explicit description of the temporal evolution of the salt concentration over time.

Applying a mass balance between a time $t$ and a time $t + {\rm d}t$ over an infinitesimal film element of length ${\rm d} z$ both in the absence of evaporation 
and with evaporation leads us to the equation for the evolution of the salt concentration $C_s$ in the film,

\begin{equation}
\left.\frac{\rm{d} C_s}{\rm{d}t}\right|_{\mathcal{R}_h,\, \mathcal{T}}
=  2j_{\mathrm{e}}\frac{C_s}{h}.
\label{eq:evol_Salt}
\end{equation}

\noindent Using the ideal gas law for water vapor in air, we express the saturated vapor mass concentration as
\begin{equation}
c_\mathrm{sat}(C_s) = c_\mathrm{sat}\frac{P_\mathrm{sat}(C_s)}{P_\mathrm{sat}(0)},
\end{equation}

\noindent where $P_\mathrm{sat}(C_s)$ is  the saturated vapor pressure at salt concentration $C_s$, and $P_\mathrm{sat}(0) = 2340$~Pa the saturated vapor pressure of pure water.
To obtain $P_\mathrm{sat}(C_s)$ at any given $C_s$, we interpolate the data of \citet{lopez-borrell_vapor_2024}.
As shown by Eq.~\ref{eq:def_grashoff},  $\mathcal{L}$ is the only variable of the Grashoff number, and we take it equal to 20~mm as in Sec. \ref{subsec:evapSansSel}.
The resulting new expression for $j_{\mathrm{e}}$ is

\begin{equation}
j_{\mathrm{e}} = a'  \hat{c}_0 \frac{\mathcal{D}c_{\mathrm{sat}}\left(\frac{P_\mathrm{sat}(C_s)}{P_\mathrm{sat}(0)}-\mathcal{R}_h\right)}{\rho_w} \frac{\mathrm{Gr}^{1 / 4}}{\mathcal{L}},\label{eq:je_model_avec_sel}
\end{equation}

\noindent with $\mathcal{D}$ and $\rho_w$ the same values as in Eq.~\ref{eq:je_model} and $a'$ the new fit parameter.
As in the previous section, we numerically solve Eq.~\ref{eq:amincissement} and Eq.~\ref{eq:evol_Salt} down to $h>5\ $ nm, with the same expression of $j_{\mathrm{d}}$ but with the expression of $j_{\mathrm{e}}$ given by Eq. \ref{eq:je_model_avec_sel} using the function \textit{odeint} from the \textit{scipy.integrate} package.
The results for $h(t)$ are displayed as dashed black lines for the different relative humidities in Fig.~\ref{fig9}~$(a)$ and the values of the fitting parameter $a'$ are given in Table \ref{tab:Summary_a}.

This new model accurately captures the experimental data for most of its duration, down to $h =100$~nm in all humidity conditions.
Note that the thinning curves with or without salt are very similar, which means that the effect of salt on evaporation is negligible down to such thicknesses.
Nevertheless, it allows us to estimate the salt concentration over time.
The variation $C_s(t)$  for each humidity condition is displayed in Fig.~\ref{fig9}~$(b)$ as coloured lines for the different relative humidities.
The black horizontal line represents the saturation of salt in water.
We identify the time $t^\star$ at which the thickness reaches $h^\star=100 \ \mathrm{nm}$.
The corresponding salt concentrations for relative humidities $\mathcal{R}_h \in [40, 60, 80, 100]~\%$ are $C_s(t^\star) \in \{150,73,58,31\}\ \mathrm{g.L^{-1}}$.
These points are displayed as coloured
crosses in Fig.~\ref{fig9}~$(b)$.
Since the model is no longer valid below these thicknesses, we cannot predict what happens in these concentrated thin films.

Below a thickness of $100$~nm, the model predicts a more rapid thinning than observed in the experiments.
At such high salt concentrations, \citet{auregan_drainage_2024} observed instabilities in the film thickness.
The thinning slowdown noticed in experimental data below $h^{\star} =100$~nm could be attributed to those instabilities, or to a complex coupling between drainage and evaporation.

Our conclusion is that neither the drainage nor the evaporation rates are measurably affected by the addition of salt, despite the strong increase in salt concentration.
The impact of salt only appears at smaller thicknesses and is threefold:
(i) the thinning rate is smaller than predicted by the model, (ii) some thickening regions are observed, (iii) the NBF is more stable.
Among these three effects, the one that controls the film rupture seems to be the stabilisation of the NBF, which we quantify in the following.
From our model, we extract the salt concentration over time and we show that it cannot explain the evolution of the thinning curves.

\subsection{Film rupture}

\begin{center}
\begin{figure}[h]
    \centering
    \includegraphics[width=1\textwidth]{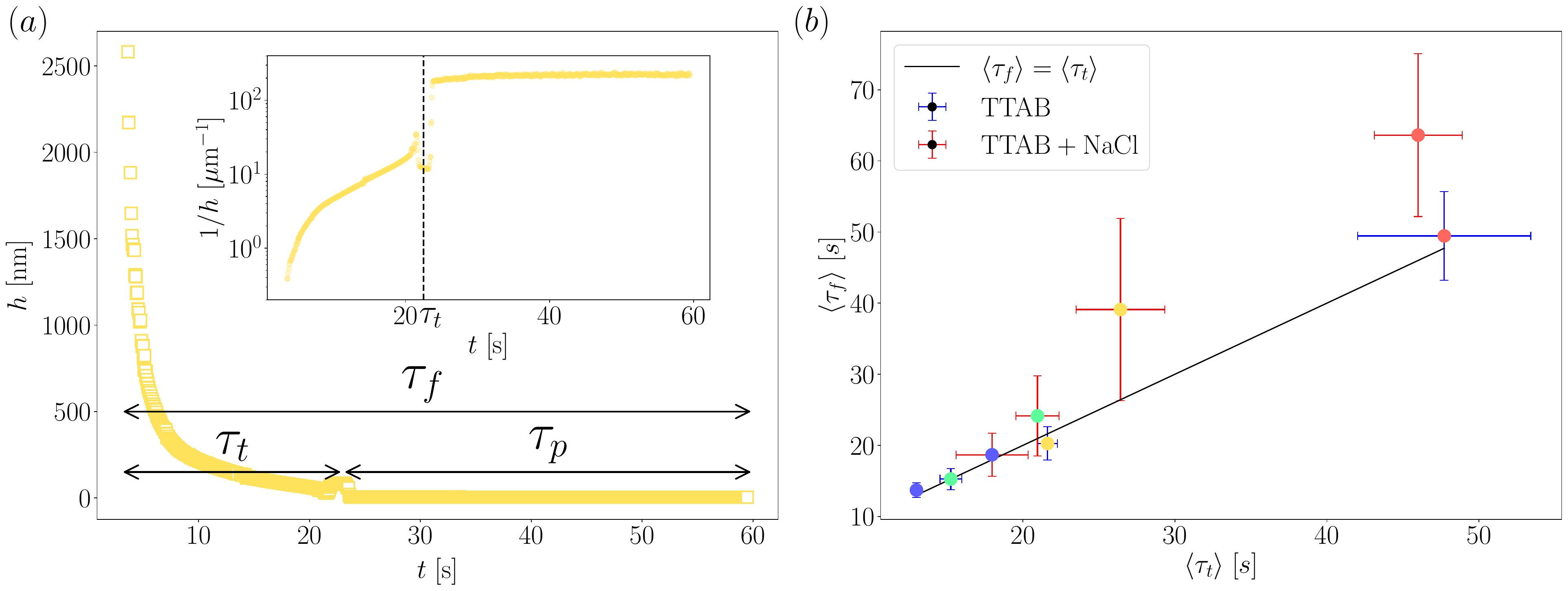}
    \caption{\textit{($a$) Illustration of the three characteristic times, the film lifetime $\tau_f$, the thinning lifetime $\tau_t$ and the plateau lifetime $\tau_p$ for a thinning curve obtained with $C_s^\circ = \ $32.5~g$.$L$^{-1}$ NaCl at a relative humidity of $\mathcal{R}_h = 80\ \%$. The inset shows the determination of $\tau_t$ from the plot of $1/h$ versus $t$ (log-lin scale). ($b$) Mean film lifetime, $\langle \tau_f \rangle$, as a function of the mean thinning time, $\langle \tau_t \rangle$, for all relative humidity conditions tested using the same colour code as previous sections, either with (\textcolor{red}{$+$}) or without (\textcolor{blue}{$+$}) 32.5~g$.$L$^{-1}$ NaCl. The solid black line corresponds to $\langle \tau_f\rangle = \langle \tau_t\rangle$.}}\label{fig5}
\end{figure}
\end{center}

As illustrated in Fig.~\ref{fig5}~$(a)$, in addition to the film lifetime $\tau_f$ determined in Fig.~\ref{fig2}, two characteristic times can be extracted from each thinning curve: the thinning lifetime $\tau_t$, and the plateau lifetime $\tau_p$.
$\tau_t$ is determined by plotting the thickness $1/h$ against $t$ as shown in the inset of Fig.~\ref{fig5}~$(a)$ in log-lin scale.
In this example, $\tau_t = 22.5$ s corresponds to the jump in $1/h$.
If the identification of this time is not as obvious as in this example, we take the intersection between a horizontal line at the mean plateau $1/h_p$ and an affine regression, $1/h = pt+q$ with $p$ and $q$ constants, fitted to the data in the range $h \in [50, 200]$~nm.
The plateau lifetime $\tau_p$ is defined as the difference $\tau_f - \tau_t$.

By extracting these two characteristic times for each set of 10 films, together with the film lifetime $\tau_f$ (Fig.~\ref{fig2}), we plot in Fig.~\ref{fig5}~$(b)$ the mean film lifetime $\langle \tau_f\rangle$ as a function of the mean thinning time $\langle \tau_t\rangle$.
The circle's colour code is the same as in Fig.~\ref{fig2} for relative humidity.
The blue and red crosses present salt-free and salt containing systems, respectively.
The $\langle \tau_f\rangle = \langle \tau_t\rangle$ line is included in black.
The collapse of all blue points onto this line indicates that film rupture occurs immediately after thinning ceases, implying that no stable plateau forms without salt.
On the other hand, the red crosses are above the $\langle \tau_f\rangle = \langle \tau_t\rangle$ line, suggesting that film rupture occurs during the plateau phase in salted systems.

Previous TFPB studies \citep{exerowa_common_1981} have shown that a Newton Black Film can be stabilised above a critical salt concentration $C_{\mathrm{el,cr}}$.
For films made with TTAB, \citet{schulze-schlarmann_disjoining_2006} reported a critical sodium chloride concentration $C_{\mathrm{el,cr}} = 23.4$~g$.$L$^{-1}$ that is lower than our initial salt concentration of $C_s^\circ= 32.5$~g$.$L$^{-1}$.
The plateau is thus probably stabilised by the same process.
A surprising feature is that no common black film is stabilised in the absence of salt.
This is unexpected as, in the absence of salt, the screening is much smaller and an electrostatic repulsion between the interfaces should appear \citep{Bergeron1997}.
The absence of CBF stabilisation could be attributed to the dynamics of the process.
However, further investigation would be necessary to clarify the mechanism.
For example, systematic measurements in a Thin Film Pressure Balance, controlling the pressure jump applied on the thin film could give insights on the transition between CBF and NBF.

\begin{center}
\begin{figure}[h]
    \centering
   \includegraphics[width=0.75\textwidth]{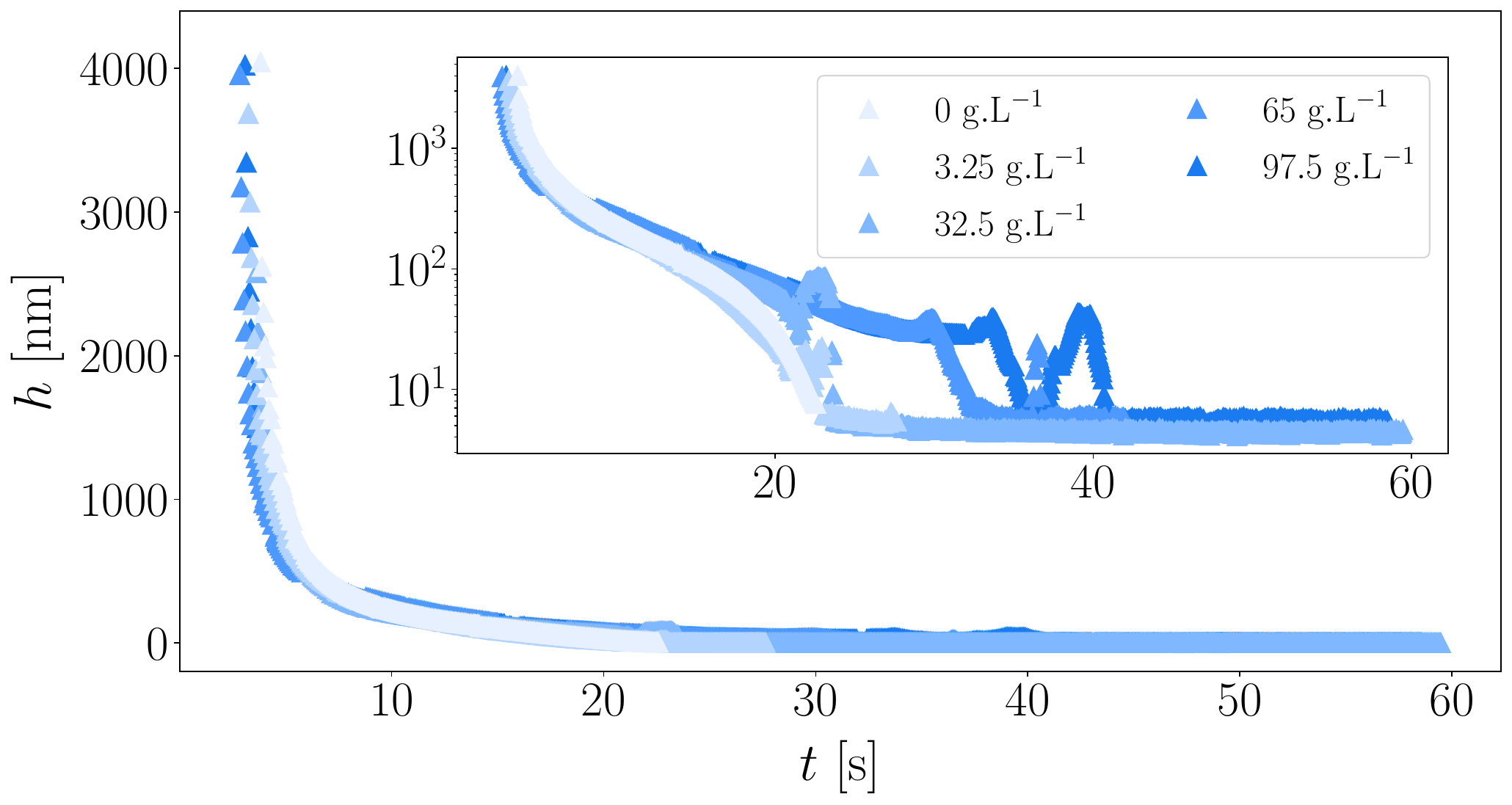}
    \caption{\textit{Thinning curves of films in Fig~\ref{fig4}~$(c)$ with the addition of the thinning curve at $\mathcal{R}_h = 80~\%$ with initial concentration $C_s^\circ \in [3.25,65,97.5]$~g$.$L$^{-1}$represented using a blue-scale gradient indicating concentration}. Insets show the data in log-lin scale.}\label{fig10}
\end{figure}
\end{center}

We performed complementary measurements at $\mathcal{R}_h = 80~\%$ with different initial concentrations $C_s^\circ$.
Fig.~\ref{fig10} shows the results of Fig.~\ref{fig4}~$(c)$ together with initial concentration $C_s^\circ \in [3.25,65,97.5]$~g$.$L$^{-1}$ represented using a blue-scale gradient indicating concentration.
We note that for $C_s^\circ \in [65,97.5]$~g$.$L$^{-1}$, most films attain the 81~mm limit of the translation stage.
Thus, in Fig.~\ref{fig10} the thinning curves do not correspond to the longer film as in Sec.~\ref{sec:FilmThickness}, but rather to the longest film attained within this limit.

\begin{center}
\begin{figure}[h]
    \centering   \includegraphics[width=0.75\textwidth]{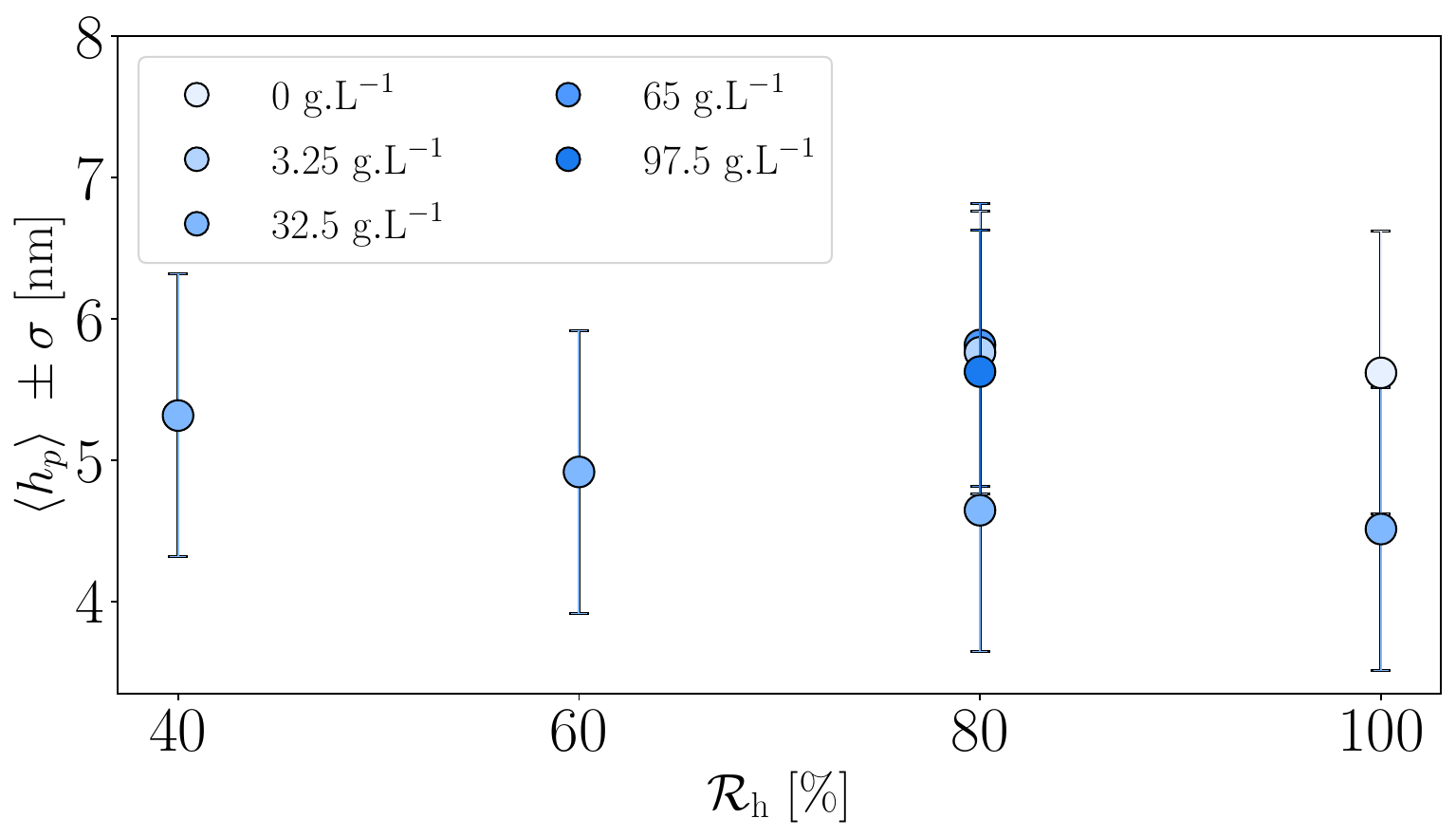}
    \caption{\textit{Mean film thickness $\langle h_p \rangle$ averaged over the plateau once local thickening is no longer observed in the thinning curves in Fig.~\ref{fig4} and Fig.~\ref{fig10}. $\sigma = 1$~nm is the uncertainty estimated on each measurement. The blue-scale gradient indicates the salt concentration.}}\label{fig11}
\end{figure}
\end{center}

The observed thickness, about 5~nm corresponds to the typical thickness of NBF as observed in the TFPB \cite{exerowa_common_1981}.
Except without added salt, an NBF forms at all concentrations, even below $C_{\mathrm{el,cr}}$.
This could be attributed to the increase of salt concentration over time, leading to a concentration higher than $C_{\mathrm{el,cr}}$ on the plateau.
In Fig.~\ref{fig11}, the thickness on the plateau $\left< h_p \right>$ measured in each experiment is plotted as a function of atmospheric humidity. $\left< h_p \right>$ is obtained by averaging the thickness on the plateau, at the later times, once local thickening is no longer observed. The uncertainty $\sigma$ is determined the error bar measured on each measure, before averaging.
We note that the thickness plateau is constant in all experiments at approximately 5$ \pm 1$~nm.

\section{Conclusion}\label{sec:Conclusion}

In this article, we investigate the influence of a high salt concentration on evaporation, drainage, and rupture of vertical soap films, pulled out of a bath containing a solution of TTAB.
We illustrate the previously reported increase in film lifetime with humidity, both with and without salt.
We also show that, at any humidity, the addition of salt leads to longer film lifetimes and a more dispersed distribution.

We discuss the stochastic nature of film rupture using the approach of \cite{shaw_dripping_1984}, plotting return maps of lifetime differences.
Surprisingly, an intermediate negative correlation was observed across all humidity conditions, regardless of whether salt was added or not.
Long films tend to be followed by shorter ones and vice versa.

The thinning curves have the same dynamics down to about 400~nm for all humidities, in the presence or in the absence of salt.
Below 400 nm, films are thicker at higher humidity.
This is interpreted as a transition from a drainage-dominated regime to a regime in which evaporation cannot be neglected.
Comparison of curves with and without salt at each relative humidity reveals that salt does not influence the thinning dynamics down to 100 nm.
The evaporation rate in this regime is quantified experimentally and compared with a model of vertical film evaporation, which accounts for the data, providing the introduction of a fitting parameter.
It demonstrates that, down to 100 nm, the effect of salt on evaporation is negligible.
Below this threshold, thinning is slowed down in the presence of salt and local thickening can appear.
These unexplained observations suggest that confinement effects may become relevant.

For films without salt, neither Common Black Films (CBF) nor Newton Black Films (NBF) are stable.
All films rupture when their thickness reaches 5-10~nm.
By contrast, in the presence of salt, a thickness plateau at 5~nm is observed at all relative humidities, indicating the systematic stabilisation of an NBF.
Relative humidity is found to positively influence the plateau lifetime.
Complementary measurements at different salt concentrations confirm the existence of the plateau, and its stability increases with salt concentration.

\section*{Acknowledgments}
Funding by the National Research Agency (ANR - 22-CE06-0029) is acknowledged.
We are grateful to Luc Deike, Giuseppe Foffi, and Luis Gómez-Nava for fruitful discussions. We also thank Laura Wallon for the measurements of the refractive indices. We acknowledge the use of LLM to improve the style of certain figures and the fluidity of certain sentences.

\section*{Declaration of interests.} The authors report no conflicts of interest.

\section*{Authors Contribution}
The study was conceived and designed by ER, AS, and VZ. Experimental execution and subsequent data analysis were carried out by VZ. The modeling was discussed and derived by ER, AS, FB and VZ. The initial manuscript was drafted by VZ, and all authors (VZ, ER, AS, FB) contributed critically to its revision and approved the final version.

\section{Surface tension of TTAB solutions with and without NaCl}\label{appA}

\begin{center}
\begin{figure}[h]
    \centering
    \includegraphics[width=0.7\textwidth]{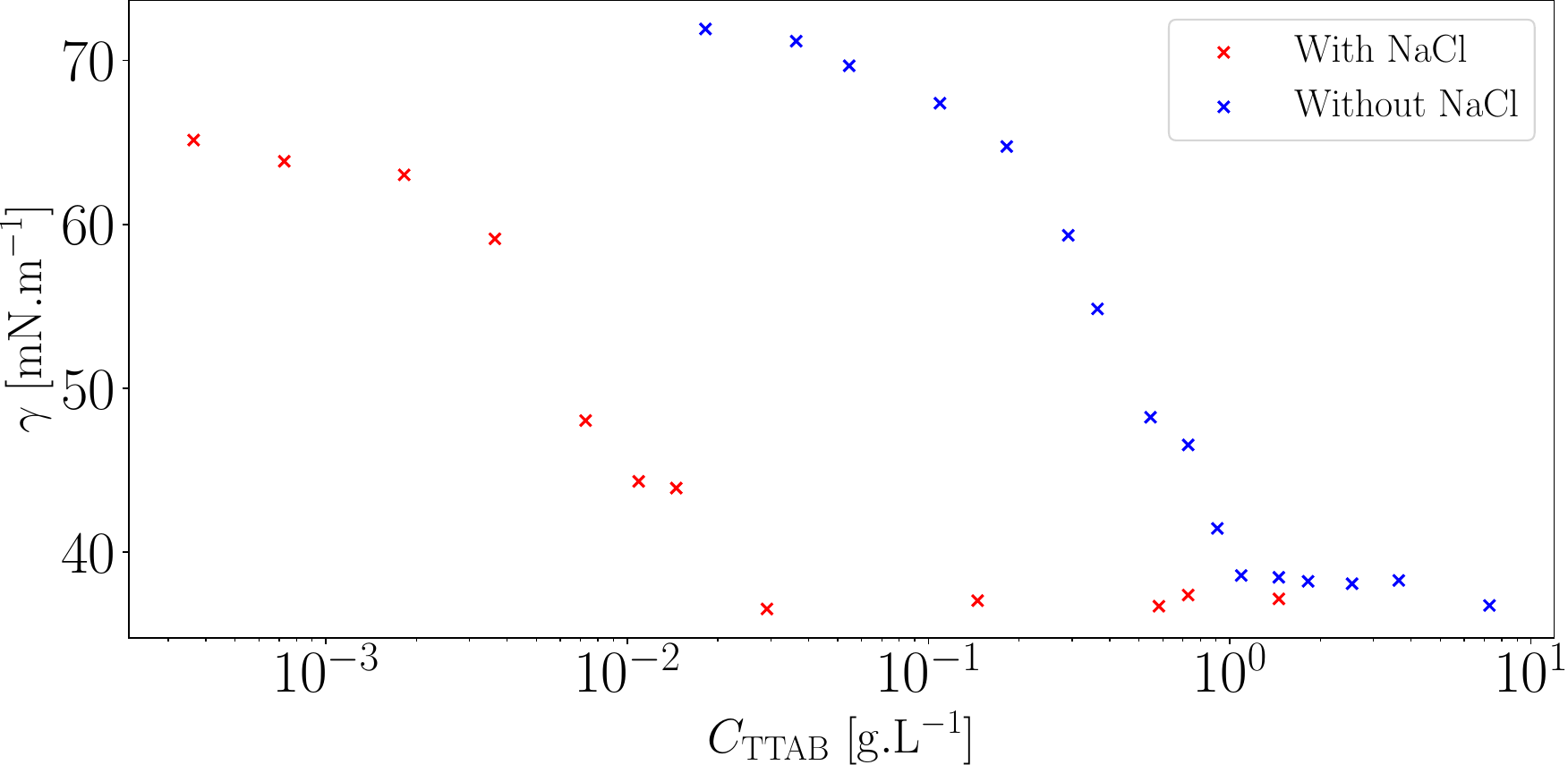}
    \caption{\textit{Surface tension as a function of TTAB concentration with (\textcolor{red}{$\times$}) and without (\textcolor{blue}{$\times$}) the addition of $ 32.5\ $g.L$^{-1}$ of NaCl.}}\label{figAppA}
\end{figure}
\end{center}

Figure \ref{figAppA} reports a surface tension measurement campaign investigating the effect of adding 32.5 g.L$^{-1}$ of NaCl to TTAB solutions.
Blue crosses (\textcolor{blue}{$\times$}) and red crosses (\textcolor{red}{$\times$}) represent the results obtained without and with NaCl, respectively.
The measurements were performed at $\mathcal{T} = 20\ ^\circ$C using a pendant drop TECLIS @TRACKER™ tensiometer.
From Fig. \ref{figAppA}, we observe a significant decrease in the critical micelle concentration, from 1.18$\ $g.L$^{-1}$ to $0.03 \ $g.L$^{-1}$.

\bibliography{ArticleJFM}

\end{document}